\documentclass[10pt,journal]{IEEEtran}
\IEEEoverridecommandlockouts
\makeatletter
\def\ps@headings{%
\def\@oddhead{\mbox{}\scriptsize\rightmark \hfil \thepage}%
\def\@evenhead{\scriptsize\thepage \hfil \leftmark\mbox{}}%
\def\@oddfoot{}%
\def\@evenfoot{}}
\makeatother
\usepackage{amssymb}
\usepackage{amsfonts,amsmath}
\usepackage[ruled]{algorithm}
\usepackage{algpseudocode}
\usepackage{algorithmicx}
\usepackage{graphicx,epsfig,subfigure}
\usepackage{subfig}
\usepackage{color}
\usepackage[table,xcdraw]{xcolor}
\usepackage{booktabs}
\usepackage{subfigure}
\usepackage{cite}

\newcommand {\mymarginpar}[1]{\marginpar{#1}}
\renewcommand {\marginpar}[1]{}

\def\_{\rule{.3em}{.15ex}}      % Get underscore by typing \_.

\newcommand{\ls}[1]
   {\dimen0=\fontdimen6\the\font
    \lineskip=#1\dimen0
    \advance\lineskip.5\fontdimen5\the\font
    \advance\lineskip-\dimen0
    \lineskiplimit=.9\lineskip
    \baselineskip=\lineskip
    \advance\baselineskip\dimen0
    \normallineskip\lineskip
    \normallineskiplimit\lineskiplimit
    \normalbaselineskip\baselineskip
    \ignorespaces
   }
\newcommand {\bearn}{\begin{eqnarray*}}
\newcommand {\eearn}{\end{eqnarray*}}
\newcommand {\barr}{\begin{array}}
\newcommand {\earr}{\end{array}}

\newcommand {\N}{{\cal N}}

\newtheorem{definition}{Definition}
\newtheorem{property}[definition]{Property}
\newtheorem{proposition}[definition]{Proposition}
\newtheorem{lemma}[definition]{Lemma}
\newtheorem{theorem}[definition]{Theorem}
\newtheorem{corollary}[definition]{Corollary}
\newtheorem{example}[definition]{Example}
\newtheorem{remark}[definition]{Remark}

\newcommand {\benum} {\begin{enumerate}}
\newcommand {\eenum} {\end{enumerate}}

\newcommand {\bdesc} {\begin{description}}
\newcommand {\edesc} {\end{description}}

\newcommand {\bfig}[2] {\begin{figure}
  \centering
  \includegraphics[width=#2]{#1}}
\newcommand {\brotatefig}[2] {\begin{figure}[htbp]
                        \centerline {
                         \epsfig{figure={#1},clip=,angle=-90,width={#2}}}}
\newcommand {\bfigfirst}[2] {\begin{figure}[h]
                        \centerline {
                        \setlength{\epsfxsize}{#2}
                        \epsffile{#1}}}
\newcommand {\efig}[2]{ \caption{#2}
                        \label{fig:#1}
                        \end{figure}
                        \mymarginpar{fig:#1}}
\newcommand {\erotatefig}[2]{ \caption{#2}
                        \label{fig:#1}
                        \end{figure}
                        \mymarginpar{fig:#1}}
\newcommand {\rfig}[1]{Figure \ref{fig:#1}}

\newcommand {\btab}[1]{
                       \begin{table}
                       \centering
                       \begin{tabular}{#1}}
\newcommand {\etab}[3] {
                       \end{tabular}
                       \caption[#3]{#2}
                       \label{tab:#1}
                       \end{table}
                       \mymarginpar{tab:#1}
                       \vspace{.1in}}

\newcommand {\btabular}[1]{\begin{center}
                       \begin{tabular}{#1}}
\newcommand {\etabular}{\end{tabular}
                       \end{center}}

\newcommand {\bdefin}[1]{\begin{definition}
                      \mymarginpar{def:#1}
                      \label{def:#1} }
\newcommand {\edefin}       {\end{definition}}

\newcommand {\bpro}[1]{\begin{property}
                      \mymarginpar{pro:#1}
                      \label{pro:#1} }
\newcommand {\epro}   {\end{property}}

\newcommand {\bprop}[1]{\begin{proposition}
                      \mymarginpar{prop:#1}
                      \label{prop:#1} }
\newcommand {\eprop}       {\end{proposition}}

\newcommand {\blem}[1]{\begin{lemma}
                      \mymarginpar{lem:#1}
                      \label{lem:#1} }
\newcommand {\elem}   {\end{lemma}}

\newcommand {\bthe}[1]{\begin{theorem}
                      \mymarginpar{the:#1}
                      \label{the:#1} }
\newcommand {\ethe}   {\end{theorem}}
\newcommand {\rthe}[1]{Theorem \ref{the:#1}}

\newcommand {\bproof}{\noindent {\bf Proof.} \ }
\newcommand {\eproof} {\hfill \squares \\ \vspace{.2cm}}
\newcommand {\bcor}[1]{\begin{corollary}
                      \mymarginpar{cor:#1}
                      \label{cor:#1} }
\newcommand {\ecor}   {\end{corollary}}

\newcommand {\bax}[1]{\begin{axiom}
                      \mymarginpar{ax:#1}
                      \label{ax:#1} }
\newcommand {\eax}       {\vspace{-.1in} \end{axiom}}

\newcommand {\bex}[2]{\vspace{.1in}
                      \begin{example}
                      \mymarginpar{ex:#1}
                       {\bf #2}
                      \label{ex:#1} }
\newcommand {\eex}       {\end{example} \vspace{.3cm} }
\newcommand {\rex}[1]{Example \ref{ex:#1}}

\newcommand {\brem}[1]{\begin{remark}
                      \mymarginpar{rem:#1}
                      \label{rem:#1} \em }
\newcommand {\erem}   {\end{remark}}

\newcommand {\beq}[1]{\mymarginpar{eq:#1}
                      \begin{equation}
                      \label{eq:#1} }

\newcommand {\beqno}[1]{\mymarginpar{eq:#1}
                      \begin{eqnarray}
                      \nonumber}

\newcommand {\eeq}       {\end{equation}}
\newcommand {\eeqno}       { && \end{eqnarray}}
\newcommand {\req}[1]{(\ref{eq:#1})}

\newcommand {\bear}[1]{\mymarginpar{eq:#1}
                       \begin{eqnarray}
                       \label{eq:#1} }

\newcommand {\bearno}[1]{\mymarginpar{eq:#1}
                       \begin{eqnarray}
                       \nonumber}

\newcommand {\eear}{\end{eqnarray}}
\newcommand {\eearno}{\end{eqnarray}}
\newcommand {\bsel}{\left \{ \begin{array}{cl}}
\newcommand {\esel}{\end{array} \right.}

\newcommand {\bmat}[1]{\left [ \begin{array}{#1}}
\newcommand {\emat}{\end{array} \right ]}
\newcommand {\bsec}[2]{\mymarginpar{sec:#2}
                       \section{#1}
                       \label{sec:#2} }

\newcommand {\rsec}[1]{Section \ref{sec:#1}}

\newcommand {\bsubsec}[2]{\mymarginpar{sec:#2}
                       \subsection{#1}
                       \label{sec:#2} }

\newcommand {\bsubsubsec}[2]{\mymarginpar{sec:#2}
                       \subsubsection{#1}
                       \label{sec:#2} }

\def\R{I\kern-0.30em R}
\def\N{I\kern-0.30em N}
\def\P{I\kern-0.30em P}
\newcommand\squares{\vrule height6pt width7pt depth1pt}

\begin{document}

\title{Efficient Multichannel Rendezvous Algorithms without Global Channel Enumeration}
%\title{Using Locality-sensitive Hashing for Rendezvous Search}

\author{Yi-Chia Cheng and
		Cheng-Shang~Chang,~\IEEEmembership{Fellow,~IEEE}\\
		\IEEEcompsocitemizethanks{\IEEEcompsocthanksitem
		The authors are with the Institute of Communications Engineering, National Tsing Hua University, Hsinchu 300044, Taiwan R.O.C. Email:  yichiacheng2001@gmail.com; cschang@ee.nthu.edu.tw.
		\protect\\
	}
	\thanks{%Manuscript received November 17, 2020; revised March 29, 2021; accepted May 16, 2021.
Part of this work has been presented in IEEE 2024 33rd Wireless and Optical Communications Conference (WOCC) \cite{wocc2024}. This work was supported in part by the National Science and Technology Council, Taiwan, under Grant 111-2221-E-007-038-MY3 and 111-2221-E-007-045-MY3. (Corresponding
author: Cheng-Shang Chang.)}}

%\markboth{IEEE Transactions on Cognitive Communications and Networking}%
%{Submitted paper}

\maketitle

\begin{abstract}
The multichannel rendezvous problem (MRP) is a critical challenge for neighbor discovery in IoT applications, requiring two users to find each other by hopping among available channels over time. This paper addresses the MRP in scenarios where a global channel enumeration system is unavailable. 
To tackle this challenge, we propose a suite of low-complexity multichannel rendezvous algorithms based on locality-sensitive hashing (LSH), tailored for environments where channel labels are unique $L$-bit identifiers rather than globally coordinated indices.
Inspired by consistent hashing techniques in distributed systems, we develop the LC-LSH and LC-LSH4 algorithms for synchronous and asynchronous settings, respectively. These algorithms significantly reduce implementation complexity while maintaining expected time-to-rendezvous (ETTR) performance comparable to state-of-the-art methods that require global channel enumeration. To ensure bounded maximum time-to-rendezvous (MTTR) in the asynchronous setting, we further introduce the ASYM-LC-LSH4 and QR-LC-LSH4 algorithms by embedding multiset-enhanced modular clock and quasi-random techniques into our framework.
Extensive simulations demonstrate that the proposed algorithms achieve performance comparable to state-of-the-art LSH algorithms in both synchronous and asynchronous settings, even without a global channel enumeration system. 
\end{abstract}


%\bsec{Introduction}{introduction}

%\IEEEPARstart{W}{ireless} networks

%\bsec{Problem statement}{statement}
\bsec{Introduction}{introduction}

In the realm of the Internet of Things (IoT), devices frequently operate in settings with numerous communication channels. A pivotal aspect for these devices is the necessity to discover one another before initiating communication, especially in dynamic environments where devices are regularly joining or exiting the network. Additionally, the presence of primary users may obstruct some communication channels. This scenario leads to the multichannel rendezvous problem (MRP), where two secondary users (IoT devices) are required to rendezvous on a commonly available channel by hopping over their available  channels with respect to time.
The MRP poses a critical challenge for neighbor discovery in many IoT applications, as discussed in various studies such as those in \cite{Theis2011,Bian2013}.

The MRP has garnered significant interest recently, highlighted in several publications including a notable book \cite{Book} on the subject. As outlined in the paper \cite{GAP2019}, the MRP involves three essential components: (i) users, (ii) time, and (iii) channels. Depending on the assumptions regarding these components, various channel hopping (CH) algorithms have been developed.

A particularly challenging setting for MRP is the {\em oblivious setting}, described in detail in the aforementioned book. In this setting, four conditions are present: (i) symmetric (sym): users are indistinguishable, (ii) asynchronous (async): there is a lack of synchronization among users' clocks, (iii) heterogeneous (hetero): different users may not share identical sets of available channels, and (iv) locally labeled (local): each user's channel labels are unique and may not align with global labels. In such a scenario, failed rendezvous attempts do not yield any useful feedback.

Research, including a study in \cite{ToN2017}, has demonstrated that the expected time-to-rendezvous (ETTR) in this setting is at least \((n_1 n_2 + 1) / (n_{1,2} + 1)\), where \(n_1\) and \(n_2\) represent the number of available channels for users 1 and 2, respectively, and \(n_{1,2}\) denotes the number of channels available to both users. Intriguingly, a random algorithm where each user independently chooses a channel from its available channel set achieves an ETTR of \(n_1 n_2 / n_{1,2}\). This performance is remarkably close to the theoretical lower bound, especially when \(n_{1,2}\) is not excessively small, suggesting that the random algorithm is nearly optimal in terms of ETTR for the sym/async/hetero/local MRP.

However, if all manufacturers of various IoT devices can agree on an international standard for labeling channels, then
a global channel enumeration system can be established. Under such an assumption, channels are globally labeled (i.e., global) from 0 to $N-1$, where $N$ is the total number of channels in a wireless network. 
Therefore, the problem falls into the category of sym/async/hetero/global MRP. 
For such an MRP, many works have focused on the maximum time-to-rendezvous (MTTR). In particular, it has been shown that the MTTR of the CH sequences proposed in \cite{Chen14,Improved2015,Chang18,gu2020heterogeneous} is $O((\log\log N)n_1 n_2)$. However, the ETTR of these CH sequences do not perform as well as that of the random algorithm.
To improve the ETTR performance, the locality-sensitive hashing (LSH) technique was first applied in \cite{LSH}. LSH, a method well-described in \cite{Leskovec2020}, is designed to hash similar items into the same bucket with a high likelihood, thereby maximizing hash collisions for similar items. This approach is particularly effective when the available channel sets of two users are similar. For the synchronous setting, the LSH algorithm \cite{LSH} demonstrated an ETTR approximating $1/J$, where $J$ is the Jaccard index between the two sets of available channels, i.e., \(J=\frac{n_{1,2}}{n_1 + n_2 - n_{1,2}}\). Notably, this ETTR is significantly lower than that achieved by a random algorithm.
In the context of the asynchronous setting, the LSH4 algorithm \cite{LSH} leverages locality-sensitive hashing  for dimensionality reduction and it achieves a lower ETTR compared to the random algorithm when a technical condition is satisfied.

The assumption that asks all manufacturers of various IoT devices to agree on an international standard for labeling channels might not be realistic. Even with such a standard, there is also a backward compatibility problem. A new generation of IoT devices might be capable of using more channels than those in the previous generation. This implies that IoT devices from different generations might have different views of the total number of channels. 

The aim of this paper is to address the challenge of the MRP without relying on a global channel enumeration system. 
As each channel in a wireless network is physically associated with a unique frequency, it can be represented by a floating-point number with a finite number of bytes. For example, if the frequency is represented using a single-precision floating-point number (4 bytes), then each channel is associated with a unique 32-bit identifier (ID).
Thus, it is reasonable to assume that each channel is represented by
a unique ID of $L$ bits.
This $L$-bit ID can, in principle, be converted into an integer, thereby creating a global channel enumeration system with a total of $N=2^L$ channels. However, for such a large $N$, the LSH algorithm \cite{LSH} and many existing methods in the literature become computationally demanding.

The main contributions of this paper are summarized as follows:

\begin{itemize}
    \item We address the MRP without relying on a global channel enumeration system by assuming that each channel is identified by a unique $L$-bit ID rather than a global index.

    \item We propose a low-complexity locality-sensitive hashing (LC-LSH) algorithm for the synchronous MRP setting. By using virtual frequencies and consistent hashing techniques, the algorithm significantly reduces implementation complexity while achieving ETTR close to state-of-the-art LSH-based methods.

    \item We extend the LC-LSH algorithm to the asynchronous setting and develop the LC-LSH4 algorithm using a dimensionality reduction approach. This algorithm achieves low ETTR even in the absence of global channel labels.

    \item To ensure a bounded maximum time-to-rendezvous (MTTR), we integrate the LC-LSH4 algorithm with existing techniques:
    \begin{itemize}
        \item We propose the ASYM-LC-LSH4 algorithm by embedding the multiset into the modular clock framework, which guarantees rendezvous within a bounded number of slots under the asymmetric setting.
        \item We propose the QR-LC-LSH4 algorithm by incorporating a quasi-random symmetrization mapping into the LC-LSH4 framework, which ensures bounded MTTR under the symmetric setting.
    \end{itemize}

    \item We conduct extensive simulations to evaluate the performance of our proposed algorithms. Results demonstrate that the ETTR and MTTR of our algorithm in the asynchronous setting, even without a global channel enumeration system, are comparable to the performance reported in \cite{LSH} with such a system.
                
\end{itemize}

The rest of the paper is organized as follows. In \rsec{mrp}, we provide a brief introduction to the multichannel rendezvous problem and review related works. In \rsec{Chashing}, we introduce low-complexity algorithms, including the LC-LSH algorithm for the synchronous setting and the LC-LSH4 algorithm for the asynchronous setting. In \rsec{qr}, we build upon the multiset-enhanced modular clock algorithm and propose the ASYM-LC-LSH4 algorithm and the QR-LC-LSH4 algorithm, which theoretically ensure an MTTR upper bound for the LC-LSH4 algorithm. In \rsec{sim}, we present simulation results to compare our algorithms with the random algorithm, the LSH algorithms in \cite{LSH}, and the quasi-random algorithm proposed by \cite{Quasi2018}. Finally, we conclude the paper in \rsec{con}.

 \bsec{The multichannel rendezvous problem}{mrp}

\bsubsec{Classification of the Problem}{class}

As discussed in \cite{GAP2019}, the multichannel rendezvous problem (MRP) involves three essential components: (i) users, (ii) time, and (iii) channels. Based on these components, MRP can be classified into various settings.

\paragraph{Users}
In the symmetric setting (\textbf{sym}), users are indistinguishable and must follow the same algorithm. In contrast, the ID setting (\textbf{ID}) assumes that users have unique identifiers and are therefore distinguishable. A special case of this is the asymmetric setting (\textbf{asym}), where users can be distinguished by one-bit IDs and assigned different roles, allowing them to use different algorithms.

\paragraph{Time}
In the synchronous setting (\textbf{sync}), the clocks of two users are synchronized to a global clock. In the asynchronous setting (\textbf{async}), the users' clocks may be unsynchronized.

\paragraph{Available Channels}
In the homogeneous setting (\textbf{homo}), both users share the same set of available channels. In the heterogeneous setting (\textbf{hetero}), the users may have different sets of available channels.

\paragraph{Labels of Channels}
In the globally labeled setting (\textbf{global}), both users assign the same labels to the same channels. In the locally labeled setting (\textbf{local}), users label channels independently, and the labels may not match. 
In this paper, we introduce a new setting called the \textbf{ID} setting, where each channel is assigned a unique $L$-bit ID.

\bigskip

As in \cite{GAP2019}, we describe the classification of an MRP using the notation:
\[
A/B/C/D,
\]
where $A$ is the abbreviation for the setting of {\em users}, $B$ is the abbreviation for the setting of {\em time}, $C$ is the abbreviation for the setting of {\em available channels}, and $D$ is the abbreviation for the setting of {\em labels of channels}.

\bsubsec{Problem Formulation}{problem}

The MRP in a wireless network is commonly referred to as the problem for two users (IoT devices) to rendezvous on a commonly available channel by hopping over their available  channels with respect to time. One common assumption made in the literature for the MRP is that there is a global channel enumeration system, i.e., channels are globally labeled from 0 to $N-1$, where $N$ is the total number of channels in a wireless network.
In this paper, we address the challenge of the MRP without a global channel enumeration system.
However, we still assume that each channel is associated with a unique frequency that can be represented by an $L$-bit identifier (ID).
Certainly, this $L$-bit ID can be converted into an integer, resulting in a global channel enumeration system where the total number of channels is $N =2^L$.

\begin{figure}[ht]
	\centering
	\includegraphics[width=0.45\textwidth]{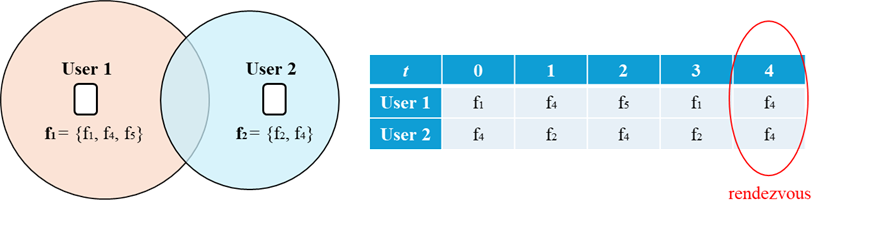}
	\caption{An illustration of the multichannel rendezvous problem.}
	\label{fig:mrpex}
\end{figure}

Specifically, the available channel set for user $i$, $i=1,2$, is represented by a set of frequencies, $${\bf f}_i=\{ f_i(0), \ldots, f_i(n_i-1)\},$$
where $n_i=|{\bf f}_i|$ is the number of available channels to user $i$.
We assume that there is at least one channel that is commonly available to the two users (as otherwise, it is impossible for the two users to rendezvous), i.e.,
\beq{avail1111}
{\bf f}_1 \cap {\bf f}_2 \ne \varnothing.
\eeq
Let $n_{1,2}=|{\bf f}_1 \cap {\bf f}_2 |$ be the number of common channels between these two users.
As shown in \rfig{mrpex}, in this case, since user 1 has the available channel set ${\bf f}_1 = \{f_1, f_4, f_5\}$, and user 2 has the available channel set ${\bf f}_2 = \{f_2, f_4\}$, the two users have one common channel, $\{f_4\}$.

As in the MRP literature, we consider the discrete-time setting, where time is divided into time slots, indexed from $t=0,1,2,\ldots$. The time-to-rendezvous (TTR) is defined as the number of time slots needed for two users to hop to a common channel.
Take \rfig{mrpex}, for example. Since the two users hop to the same common channel $f_4$ when $t=4$, the TTR is therefore 5.

\bsubsec{Related Work}{relatedmrp}

\iffalse
In the literature, there are two basic methods for solving an MRP: (i) methods that use the algebraic structure of various combinatorial designs, and (ii) methods that break the symmetry of the two users by using IDs.
There are several well-known {\em combinatorial design} algorithms, including CRSEQ \cite{CRSEQ}, JS \cite{JS2011}, DRDS \cite{DRDS13}, T-CH \cite{Matrix2015}, DSCR \cite{DSCR2016}, IDEAL-CH \cite{GAP2019}, RDSML-CH \cite{Wang2022}.
The length of these CH sequences is $O(N^2)$ and thus difficult to implement for a very large $N$.
On the other hand, there are several {\em symmetry-breaking} algorithms in the literature, see e.g., 
\cite{Chen14,Improved2015,Chang18,gu2020heterogeneous} with $O((\log\log N)n_1 n_2)$ MTTR
and \cite{Quasi2018,QECH} with $O((\log N)n_1 n_2)$ MTTR. Among these symmetry-breaking algorithms,
two of them can be used for a large $N$: (i) FMR \cite{Chang18} that uses the complete symmetrization mapping for symmetry-breaking,
and (ii) QR \cite{Quasi2018} that uses the strong ternary symmetrization mapping for symmetry-breaking.
The MTTR of FMR is $O((\log\log N)n_1 n_2)$, but its ETTR is much larger than that of the random algorithm (as redundancy is added into the algorithm to guarantee the rendezvous). On the other hand,
the MTTR of QR is $O((\log N)n_1 n_2)$ and its ETTR is almost the same as that of the random algorithm.
\fi

In the literature, two main approaches have been developed to solve the MRP:  
(i) methods based on the algebraic structure of various combinatorial designs, and  
(ii) methods that break the symmetry between the two users using IDs.
Several well-known \emph{combinatorial design} algorithms include CRSEQ~\cite{CRSEQ}, JS~\cite{JS2011}, DRDS~\cite{DRDS13}, T-CH~\cite{Matrix2015}, DSCR~\cite{DSCR2016}, IDEAL-CH~\cite{GAP2019}, and RDSML-CH~\cite{Wang2022}. However, these algorithms typically produce CH sequences of length \(O(N^2)\), which makes them impractical to implement when \(N\) is very large.
On the other hand, several \emph{symmetry-breaking} algorithms have been proposed, including those in~\cite{Chen14,Improved2015,Chang18,gu2020heterogeneous} with MTTR of \(O((\log \log N) n_1 n_2)\), and those in~\cite{Quasi2018,QECH} with MTTR of \(O((\log N) n_1 n_2)\). Among these, two algorithms are suitable for large values of \(N\):  
(i) the FMR algorithm~\cite{Chang18}, which employs complete symmetrization mapping; and  
(ii) the QR algorithm~\cite{Quasi2018}, which uses strong ternary symmetrization mapping.
While the MTTR of FMR is \(O((\log \log N) n_1 n_2)\), its ETTR is significantly higher than that of the random algorithm due to added redundancy for guaranteed rendezvous. In contrast, the QR algorithm achieves an MTTR of \(O((\log N) n_1 n_2)\) and offers ETTR performance nearly equivalent to that of the random algorithm.

Our work in this paper is primarily inspired by the locality-sensitive hashing (LSH) algorithms in
 \cite{LSH}, in particular LSH2 and LSH4 for globally labeled channels.  For the synchronous setting, LSH2 uses two pseudo-random permutations $\pi_1$ and $\pi_2$ to generate the channel hopping sequence $\{c(t), t=1,2, \ldots, N\}$
for an available channel set ${\bf c}=\{c_0, c_1, \ldots, c_{n-1}\}$ by selecting
$c(t)=c_{i^*}$, where  $i^*$ is the index of the channel that minimizes the modular difference between
$\pi_1(c_i^*)$ and $\pi_2(t)$,
i.e.,
\beq{LSH2}
i^*={\rm argmin}_{0 \le i \le n-1}((\pi_1(c_i)-\pi_2(t))\;\mod\;N).
\eeq
For the asynchronous setting,  LSH4 adopts a novel dimensionality reduction technique, mapping the available channel set of a user to a much smaller multiset to increase the rendezvous probability.

In this paper, we use the random algorithm, the LSH algorithms in \cite{LSH}, and the QR algorithm proposed in \cite{Quasi2018} as our benchmarks.

\section{The proposed low-complexity locality-sensitive hashing algorithms}
\label{sec:Chashing}

\bsubsec{The sym/sync/hetero/ID MRP}{sync}

In this section, we consider the sym/sync/hetero/ID MRP, where (i) the two users are indistinguishable, (ii) their clocks are synchronized, (iii) their available channel sets may differ, and (iv) each user's channel labels are uniquely identified by an $L$-bit ID.

As discussed in \rsec{introduction}, an $L$-bit ID can be converted into an integer, resulting in a global channel enumeration system with a total of $N = 2^L$ channels. However, for $L = 32$, implementing channel hopping algorithms over $N = 2^{32}$ channels is computationally infeasible for most existing methods in the literature. For example, as illustrated in \rsec{mrp}, the LSH2 algorithm in \cite{LSH} relies on two pseudo-random permutations of the $N$ channels to redistribute them. Generating pseudo-random permutations over such a large domain ($N = 2^{32}$) is highly computationally demanding.
To address this challenge, we propose a family of low-complexity locality-sensitive hashing (LC-LSH) algorithms that significantly reduce the computational cost while maintaining high rendezvous efficiency.

\iffalse
In this section, we consider the sym/sync/hetero/ID MRP, where (i) the two users are indistinguishable, (ii) the clocks of the two users are synchronized, (iii) the two users may not share identical sets of available channels, and (iv) each user's channel labels are uniquely identified by an $L$-bit ID.
As mentioned in \rsec{introduction}, one can convert the $L$-bit ID into an integer to obtain a global channel enumeration system where the total number of channels is $N=2^L$. However, for $L=32$, implementing channel hopping algorithms  of $N=2^{32}$ channels is extremely difficult for most existing channel hopping algorithms in the literature. For example, 
as depcited in \rsec{mrp}, the LSH2 algorithms in \cite{LSH} use two pseudo-random permutations of the $N$ channels to redistribute the $N$ channels. Implementing a pseudo-random permutation of $N=2^{32}$ channels is computationally difficult. Therefore, we propose our low-complexity locality-sensitive hashing (LC-LSH) algorithms to reduce the computational cost.
\fi

%\bsubsec{The synchronous setting}{sync}

The core idea of our LC-LSH algorithm is to expand the hash space from the interval \([0, N - 1]\) to the interval \([0, KN - 1]\), and to map each frequency into this expanded space \(K\) times. As a result, each frequency is associated with \(K\) \emph{virtual frequencies}. When \(K\) is large, the law of large numbers implies that the intervals corresponding to the available channels become nearly evenly spaced. This approach eliminates the need to redistribute available channels using pseudo-random permutations, as done in the LSH2 algorithm in \cite{LSH}.

For the ease of our presentation, we simply assume that $K$ is a power of 2, i.e., $\log_2 K$ is an integer. Our algorithm consists of the following steps: (i) Map each frequency to $K$ virtual frequencies with each virtual frequency represented by an $(L+\log_2 K)$-bit ID, (ii) Hash each virtual frequency into a ring with $K 2^L$ nodes through a random permutation $\pi$ of the $(L+\log_2 K)$-bit ID, (iii) Sort the hashed values in increasing order, (iv) Linearize the ring into the interval $[0, K 2^L]$ by breaking the ring at the origin, and (v) Use the binary search for locality-sensitive hashing of a sequence of pseudo-random  numbers for channel selection at each time slot.
The detailed steps of the implementation of our {\color{black}LC}-LSH CH algorithm are shown in Algorithm \ref{alg:LCLSH}.

Note that the computational complexity of Step 1 and Step 2 is linear in the number of available channels $n$, the number of virtual frequencies $K$, and the number of bits for the binary representation $(L+\log_2 K)$.
The sorting of the $nK$ hashed values in Step 3 can be completed with $O(nK \log (nK))$.
The linearization step is $O(1)$. Finally, the binary search for channel selection in Step 5 is $O(\log (nK))$ for each time $t$.

\begin{algorithm}\caption{The low-complexity locality-sensitive hashing (LC-LSH) CH algorithm}\label{alg:LCLSH}
	
	\noindent {\bf Input}:  A set of available channels ${\bf f}=\{f_0, f_1, \ldots, f_{n-1}\}$ with each frequency represented by an $L$-bit ID, a pseudo-random permutation $\pi$  of  $(L+\log_2 K)$ elements,
	and a sequence of pseudo-random uniform numbers $\{U(t), t \ge 0\}$ in $[0, K2^L -1]$.
	
	\noindent {\bf Output}: A CH sequence $\{f(t), t \ge 0\}$ with $f(t) \in {\bf f}$.
	
	\noindent 1: {\bf (Virtual frequencies)} For each frequency $f_i$, enlarge it into $K$ virtual frequencies, indexed from $f_{i,k}$, $k=0,1,\ldots, K-1$. The $k^{th}$ virtual frequency of $f_i$ has the $(L+\log_2 K)$-bit binary representation obtained
	by appending the $\log_2 K$-bit binary representation of $k$ to the $L$-bit representation of the frequency.
	
	\noindent 2: {\bf (Hashing)} For each virtual frequency $f_{i,k}$, we obtain another  $(L+\log_2 K)$-bit binary representation  by  permuting the original $(L+\log_2 K)$-bit binary representation (in Step 1) with the pseudo-random permutation $\pi$. The hashed value of $f_{i,k}$, denoted by $h(f_{i,k})$, is obtained by converting this (permuted)
	$(L+\log_2 K)$-bit binary representation into an integer in $[0, K2^L -1]$.
	
	\noindent 3: {\bf (Sorting)} Sort the hashed values $h(f_{i,k})$, $i=0,1,\ldots, n-1$, $k=0,1, \ldots, K-1$ in increasing order.
	
	\noindent 4: {\bf (Linearization)} Suppose that the smallest hashed value is $h(f_{i^\prime,k^\prime})$ for some $i^\prime$ and $k^\prime$. Add a virtual frequency to $f_i$ with the hashed value $K2^L$. Now we have a partition of the interval $[0, K 2^L]$ by these hashed values.
	
	\noindent 5: {\bf (Binary search for channel selection)} For each time slot $t$, use the binary search to find the smallest hashed value, denoted by $h(f_{i^*,k^*})$, that is not less than $U(t)$. At time $t$, select $f_{i^*}$, i.e., $f(t)=f_{i^*}$.
	
\end{algorithm}

\begin{figure}[ht]
	\centering
	\includegraphics[width=0.45\textwidth]{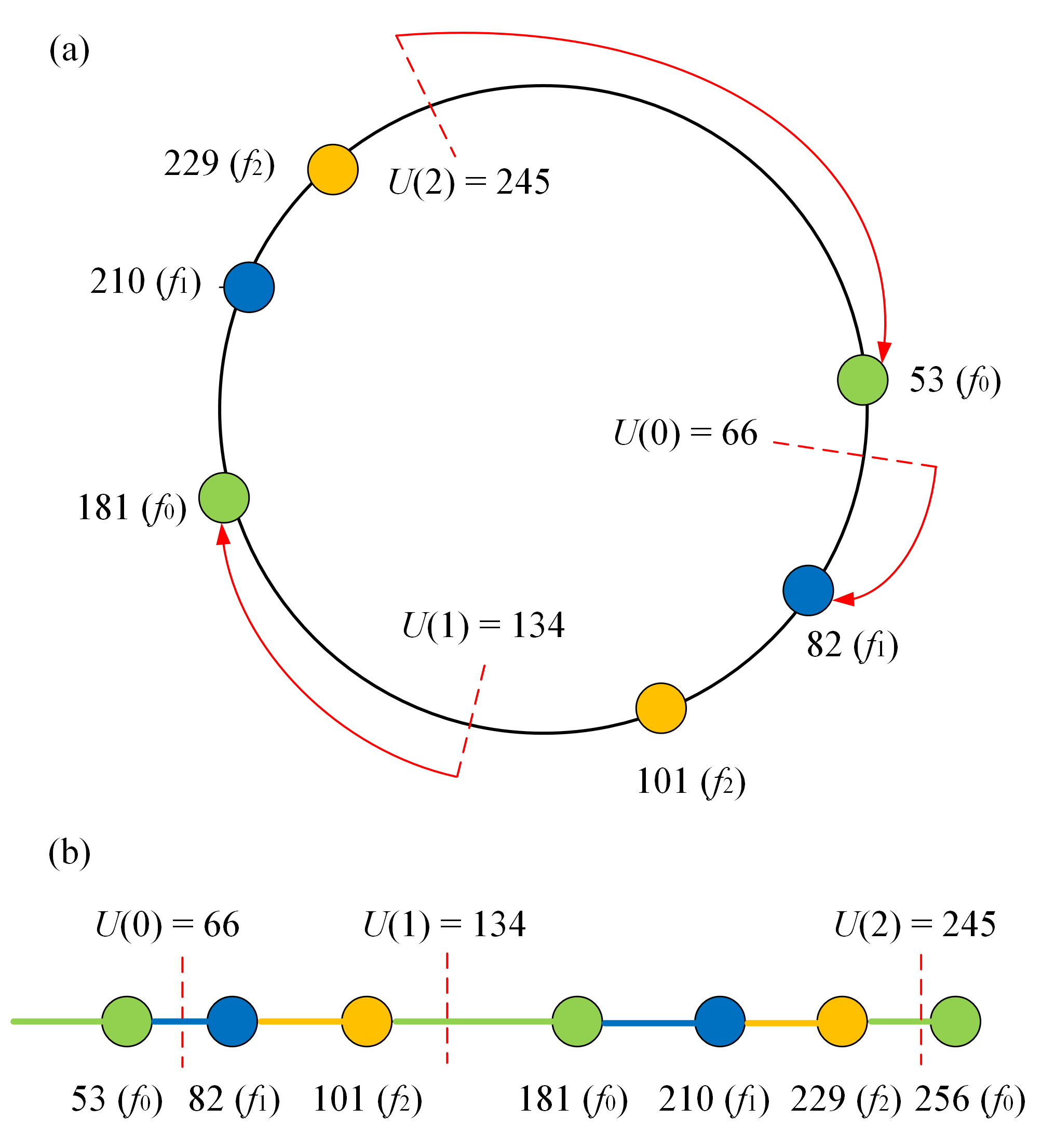}
	\caption{An illustration of the LC-LSH CH algorithm in \rex{LCLSH}: (a) hashing to the ring, (b) linearization for binary search.}
	\label{fig:LCLSH}
\end{figure}

\bex{LCLSH}{}
To illustrate how the {\color{black}LC}-LSH algorithm works, consider a user with three frequencies $\{f_0,f_1,f_2\}$. These three frequencies have the 7-bit binary representations: 
$$\{0110101, 1010010, 1100101\}.$$ Thus, we have $n=3$ and $L=7$.
Suppose we choose $K=2$ and map each frequency to two virtual frequencies. Then
the two virtual frequencies $f_{0,0}$ and $f_{0,1}$ corresponding to $f_0$ have the 8-bit binary representation $\{0110101{\bf 0}, 0110101{\bf 1}\}$.
Similarly, $f_{1,0}$ and $f_{1,1}$ have the 8-bit binary representation $\{10100100, 10100101\}$, and
$f_{2,0}$ and $f_{2,1}$ have the 8-bit binary representation $\{11001010, 11001011\}$.
For the ease of presentation, suppose that we choose the circular
permutation $\pi: (01234567) \mapsto (70123456)$
to hash these virtual frequencies into $[0,255]$. The hashed value
$h(f_{0,0})$ is the integer with binary representation $00110101$, i.e., $h(f_{0,0})=53$.
Similarly, $h(f_{0,1})$ is the integer associated with binary representation $10110101$, i.e., $h(f_{0,1})=181$.
Also, $h(f_{1,0})=82$, $h(f_{1,1})=210$, $h(f_{2,0})=101$ and $h(f_{2,1})=229$. See \rfig{LCLSH}(a) for an illustration of hashing to the ring ranging from 0 to 255. Now we sort these virtual frequencies in increasing order and this yields $\{53(f_0), 82(f_1), 101(f_2), 181(f_0), 210(f_1), 229(f_2)\}$.
For linearization, we add a virtual frequency associated with $f_0$ at 256 to form a partition of the interval $[0,256]$ into the union of the following disjoint intervals
$[0,53] (f_0)$, $[54, 82] (f_1)$, $[83,101] (f_2)$, $[102,181] (f_0)$, $[182,210] (f_1)$, $[211,229] (f_2)$, and $[230,256] (f_0)$.
See \rfig{LCLSH}(b) for an illustration of linearization. The linearization step converts the ring into an interval $[0,256]$ so that the binary search can be used. Now suppose $U(0)=66$, $U(1)=134$ and $U(2)=245$. Since $U(0)$ is in the interval $[54, 82]$,
we choose $f(0)=f_1$. Similarly, $f(1)=f_0$ and $f(2)=f_0$.
\eex

Compared to LSH2 in \cite{LSH}, the online computational complexity of our LC-LSH CH algorithm in Algorithm \ref{alg:LCLSH} is greatly reduced from $O(n)$ for linear search to $O(\log (Kn))$ for binary search. Moreover, the offline computational complexity is also greatly reduced from implementing a random permutation of $N=2^L$ elements to a random permutation of $\log_2(K)+L$ elements. In Theorem 1 \cite{LSH}, it was shown that the ETTR of the LSH2 algorithm approaches $1/J$ when $N \to \infty$, where $J=\frac{n_{1,2}}{n_1+n_2-n_{1,2}}$ represents the Jaccard index of the available channel sets of the two users. It is expected that the ETTR of the LC-LSH algorithm will also approach $1/J$. This will be verified by simulations in \rsec{sim}.

\bsubsec{The sym/async/hetero/ID MRP}{async}

In this section, we consider the sym/async/hetero/ID MRP, where (i) the two users are indistinguishable, (ii) their clocks may not be synchronized, (iii) their available channel sets may differ, and (iv) each user's channel labels are uniquely identified by an $L$-bit ID.

In constructing the CH sequence in this setting, our approach mirrors the concept of dimensionality reduction akin to that in LSH4 \cite{LSH}. Initially, the LC-LSH algorithm in Algorithm \ref{alg:LCLSH} is employed to create a CH sequence $\tilde f(t)$ over the intervals $t=0,1,\ldots, T_0-1$. This process gives rise to the multiset $\tilde {\bf f}_i=[\tilde f_i(0), \tilde f_i(1), \ldots, \tilde f_i(T_0-1)]$ for each user $i=1$ and 2.
Since there is a non-zero chance of the multisets' intersection being null, a mixed strategy is adopted in LSH4 \cite{LSH}. With a probability $p_0$ (resp. $(1-p_0)$), a user selects a channel from their multiset (resp. from the entire set of available channels). This leads to the {\color{black}LC}-LSH4 algorithm in Algorithm \ref{alg:LCLSH4}.
The insight here is that the LC-LSH algorithm generates coupled CH sequences for the two users by utilizing global channel labels, enabling the focal strategy \cite{Alpern95} to accelerate the rendezvous process.
As the LC-LSH4 algorithm is a low-complexity implementation of LSH4 \cite{LSH}, one expects that the ETTR of the LC-LSH4 algorithm is close to that of LSH4 \cite{LSH}, which is approximately
\beq{LSH41111lc}
\frac{1}{(1-p_0^2) \frac{n_{1,2}}{n_1 n_2}+ p_0^2 \frac{J}{T_0}}.
\eeq
This will be verified by simulations in \rsec{sim}.

\begin{algorithm}\caption{The LC-LSH4 CH algorithm}\label{alg:LCLSH4}
	
	\noindent {\bf Input}:  A set of available channels ${\bf f}=\{f_0, f_1, \ldots, f_{n-1}\}$  with each frequency represented by an $L$-bit ID, a pseudo-random permutation $\pi$  of  $(L+\log_2 K)$ elements,
	a sequence of pseudo-random  uniform  numbers $\{U(t), t \ge 0\}$ in $[0, K2^L -1]$, and two parameters $T_0$ and $p_0$.
	
	\noindent {\bf Output}: A CH sequence $\{f(t), t\ge 0 \}$ with $f(t) \in {\bf f}$.

	\noindent 1: Use the LC-LSH algorithm in Algorithm \ref{alg:LCLSH} to generate a CH sequence $\tilde f(t)$ for $t=0,1,\ldots, T_0-1$.
	Define the multiset $\tilde {\bf f}=[\tilde f(0), \tilde f(1), \ldots, \tilde f(T_0-1)]$.
	
	\noindent 2:  With probability $p_0$ (resp. $(1-p_0)$), randomly select a frequency $f$ from $\tilde {\bf f}$
	(resp. ${\bf f}$) and let $f(t)=f$.

\end{algorithm}

Compared to previous algorithms, our low-complexity algorithms utilize virtual frequencies (channels) to achieve a more even distribution of available channels and significantly reduce computational costs. However, the worst-case scenario still needs to be addressed. This issue is tackled by the ASYM-LC-LSH4 algorithm and the QR-LC-LSH4 algorithm, which will be discussed in the next section.

\section{Achieving bounded MTTR}
\label{sec:qr}

Our LC-LSH4 algorithm can decrease the ETTR by using the dimensionality reduction technique. However, if there are no common channels in the multiset, two users need to rely on the random algorithm to rendezvous, which means that the MTTR of the LC-LSH4 algorithm cannot be bounded. To tackle this problem, we propose the ASYM-LC-LSH4 algorithm
for the asym/async/hetero/ID MRP in \rsec{asylclsh4} and the QR-LC-LSH4 algorithm for the sym/async/hetero/ID MRP in \rsec{qrlsh4}.

\bsubsec{The asym/async/hetero/ID MRP}{asylclsh4}

In this section, we consider the asym/async/hetero/ID MRP, where (i) the two users are assigned two different roles, (ii) their clocks may not be synchronized, (iii) their available channel sets may differ, and (iv) each user's channel labels are uniquely identified by an $L$-bit ID.

Achieving a bounded MTTR is relatively easy as the two users can adopt different strategies to generate their CH sequences. One widely used approach is the modular clock algorithm proposed in \cite{Theis2011}. In this algorithm, user 1 (resp. user 2) selects a period $P_1 \ge n_1$ (resp. $P_2 \ge n_2$) to cycle through its available channels. As long as $P_1$ and $P_2$ are coprime, it then follows from the Chinese Remainder Theorem (see e.g., Theorem 4 of \cite{Theis2011})
that these two users  will rendezvous within $P_{1}P_{2}$ time slots. To ensure that $P_1$ and $P_2$ are coprime,
we index the set of primes (starting from 3) and partition them into two disjoint sets: the odd-indexed set  ${\bf P}_{\rm odd}$ and the even-indexed ${\bf P}_{\rm even}$. Since the two users can play different roles, 
user 1 (resp. user 2) selects its period from ${\bf P}_{\rm odd}$ (resp. ${\bf P}_{\rm even}$). By construction, the selected primes are distinct and hence coprime. This approach guarantees a bounded MTTR even when the available channels are locally labeled.  However, its ETTR is only $n_1n_2/n_{1,2}$ even when $P_1=n_1$ and $P_2=n_2$,  which is approximately the same as that of the random algorithm.

For the asym/async/hetero/ID MRP, we know that each channel can be represented by an $L$-bit ID and we can use that information to improve the ETTR performance, as demonstrated by the LC-LSH4 algorithm in \rsec{async}. Our idea is to embed the multiset
into the modular clock algorithm \cite{Theis2011}. To this end, we propose the multiset-enhanced modular clock algorithm. 
As in LC-LSH4, we assume that the multiset $\tilde {\bf f}=[\tilde f(0), \tilde f(1), \ldots, \tilde f(T_0-1)]$ has been already generated.
In addition to the available channel set and the multiset,
the multiset-enhanced modular clock algorithm requires three parameters: the period $P$ that is an integer not smaller than the number of available channels $n$, the slope $r$ that is relatively prime to $P$, and the bias $b$ that is an integer selected from $\{0,1, \ldots, P-1\}$. For each time $t$, a clock value $k$ is computed as
$((r*t+b)\;\mbox{mod}\;P)$. If the clock value $k$ is not greater than $n-1$, then the $k^{th}$ available channel is selected. Otherwise,
 a channel is selected at random from the multiset. The detailed steps of the algorithm are shown in Algorithm \ref{alg:clock}. 

\begin{algorithm}\caption{The multiset-enhanced modular clock algorithm}\label{alg:clock}
	
	\noindent {\bf Input}: An available channel set ${\bf f}=\{f_0, f_1, \ldots, f_{n-1}\}$, a multiset $\tilde {\bf f}=[\tilde f(0), \tilde f(1), \ldots, \tilde f(T_0-1)]$, a period $P \ge n$, a slope $r>0$ that is relatively prime to $P$, a bias $0 \le b \le P-1$, and an index of time $t$.

	\noindent {\bf Output}: A CH sequence $\{f(t), t\ge 0 \}$ with $f(t) \in {\bf f}\cup \tilde {\bf f}$.
	
	\noindent 1: For each $t$, let $k=((r*t+b)\;\mbox{mod}\;P)$.
	
	\noindent 2: If $k\le n-1$, let $f(t)=f_k$.
	
	\noindent 3: Otherwise, select $f(t)$ uniformly at random  from the multiset $\tilde {\bf f}=[\tilde f(0), \tilde f(1), \ldots, \tilde f(T_0-1)]$.
	
\end{algorithm}

Suppose the two users employ the multiset-enhanced modular clock algorithm in Algorithm~\ref{alg:clock} to generate their CH sequences. We make the following two key observations:

\begin{description}
    \item[(i)] MTTR bound: Since the multiset-enhanced modular clock algorithm includes all channels from the available channel set within its period, the two users are guaranteed to rendezvous within \(P_1 P_2\) time slots if \(P_1\) and \(P_2\) are coprime.
    
    \item[(ii)] Comparable ETTR to LC-LSH4: If the period \(P\) is chosen such that \(P \approx n/(1 - p_0)\), then the probability of selecting a channel from the multiset is approximately \(1 - n/P \approx p_0\). Under this condition, the ETTR is expected to be comparable to that of LC-LSH4.
\end{description}

Motivated by these observations, we propose the asymmetric LC-LSH4 (ASYM-LC-LSH4) algorithm, as detailed in Algorithm~\ref{alg:asylclsh4}. In ASYM-LC-LSH4, the two users play different roles and select distinct primes as their periods, ensuring a bounded MTTR (see \rthe{asyMTTR} for a formal proof). The algorithm also adopts the split-selection strategy from the multiset and the available channel set, as used in LC-LSH4, thereby achieving ETTR performance comparable to LC-LSH4.

\begin{algorithm}\caption{The ASYM-LC-LSH4 CH algorithm}\label{alg:asylclsh4}
	
	\noindent {\bf Input}: A set of available channels ${\bf f}=\{f_0, f_1, \ldots, f_{n-1}\}$, with each frequency represented by an $L$-bit ID, a pseudo-random permutation $\pi$ of $(L+\log_2 K)$ elements, a sequence of pseudo-random uniform numbers $\{U(t), t \ge 0\}$ in $[0, K2^L -1]$, two parameters $T_0$ and $p_0${\color{black}, two disjoint sets of primes ${\bf P}_{\rm odd}$ and  ${\bf P}_{\rm even}$, and two roles (i.e., role 1 and role 2)}.

	\noindent {\bf Output}: A CH sequence $\{f(t), t\ge 0 \}$ with $f(t) \in {\bf f}\cup \tilde {\bf f}$.
	
	\noindent 1: Use the LC-LSH algorithm in Algorithm \ref{alg:LCLSH} to generate a CH sequence $\tilde f(t)$ for $t=0,1,\ldots, T_0-1$.
	Define the multiset $\tilde {\bf f}=[\tilde f(0), \tilde f(1), \ldots, \tilde f(T_0-1)]$.
	
%	\noindent 2: Select the odd-indexed primes and group them into an odd-indexed set, i.e., ${\bf P_{\text{odd}}} = \{P_1, P_3, \ldots\}$. Similarly, select the even-indexed primes and group them into an even-indexed set, i.e., ${\bf P_{\text{even}}} = \{P_2, P_4, \ldots\}$.
	
	\noindent 2: Select the smallest prime $P$ such that $P \geq \lceil n / (1 - p_0) \rceil$ from ${\bf P}_{\rm odd}$ for role 1 (resp. from ${\bf P}_{\rm even}$ for role 2).
	
	\noindent 3: Randomly select the slope $r$ from $[1, P - 1]$, and set the bias $b = 0$.
	
	\noindent 4: Let $f(t)$ be the output channel at time $t$ from the multiset-enhanced modular clock algorithm in Algorithm \ref{alg:clock}, with the period $P$, the slope $r$, the bias $b$, and the multiset $\tilde{f}$.

\end{algorithm}

\bthe{asyMTTR}
(The MTTR Bound) 
Consider the asym/async/hetero/ID MRP. Suppose two users employ the ASYM-LC-LSH4 algorithm as described in Algorithm \ref{alg:asylclsh4}. These two users will rendezvous within $9n_1n_2 \left(\frac{1}{1-p_0}\right)^2$ time slots, where $n_1$ denotes the number of available channels for user 1, $n_2$ denotes the number of available channels for user 2, and $p_0$ is the parameter defined in the LC-LSH4 algorithm in Algorithm \ref{alg:LCLSH4}.
\ethe

\bproof
For any integer $n$, the smallest prime greater than or equal to $n$ can be found in the interval $[n, 2n]$ {\color{black}\cite{prime}}. Similarly, the second smallest prime greater than or equal to $n$ can be located in the interval $[n, 3n]$. 
Thus, the ASYM-LC-LSH4 algorithm guarantees a bounded MTTR within $9n_1n_2 \left(\frac{1}{1-p_0}\right)^2$.
\eproof

\bsubsec{The sym/async/hetero/ID MRP}{qrlsh4}

In this section, we consider the sym/async/hetero/ID MRP, where (i) the two users are indistinguishable, (ii) their clocks may not be synchronized, (iii) their available channel sets may differ, and (iv) each user's channel labels are uniquely identified by an $L$-bit ID.

%\bsubsec{The QR-LC-LSH4 algorithm}{qrlsh4}
%{\color{black}In this section, we consider the sym/async/hetero/ID MRP.}

For such an MRP, we introduce the QR-LC-LSH4 channel hopping algorithm, presented in Algorithm~\ref{alg:QR}, which ensures a bounded MTTR while maintaining ETTR performance similar to that of LC-LSH4.
The design follows the general framework of the QR algorithm proposed in \cite{Quasi2018}. Specifically, it uses the $L$-bit ID of the first channel in the multiset as the user's ID. This enables the use of the construction from \cite{ToN2017}, which generates CH sequences from binary IDs. However, the assumption in \cite{ToN2017} that user IDs are unique may not hold here, as two users could independently select the same common channel as their ID.
To overcome this limitation, the $L$-bit ID is mapped to an $M$-trit codeword $(w(0), w(1), \ldots, w(M-1))$ 
(over the alphabet \(\{0,1,2\}\)) by the 4B5B  strong ternary symmetrization mapping in \cite{Quasi2018}, where $M=\lceil L/4 \rceil *5+6$. The symbol ``2'' serves as a special case--when encountered, the user simply remains on its ID channel. This mechanism ensures that users with identical IDs can still rendezvous on the corresponding channel.

\begin{algorithm}\caption{The QR-LC-LSH4 CH algorithm}\label{alg:QR}
	
	\noindent {\bf Input}: A set of available channels ${\bf f}=\{f_0, f_1, \ldots, f_{n-1}\}$  with each frequency represented by an $L$-bit ID, a pseudo-random permutation $\pi$ of $(L+\log_2 K)$ elements, a sequence of pseudo-random uniform numbers $\{U(t), t \ge 0\}$ in $[0, K2^L -1]$, and two parameters $T_0$ and $p_0$.
	
	\noindent {\bf Output}: A CH sequence $\{f(t), t\ge 0 \}$ with $f(t) \in {\bf f}\cup \tilde {\bf f}$.
	
	\noindent 1: Use the {\color{black}LC}-LSH algorithm in Algorithm \ref{alg:LCLSH} to generate a CH sequence $\tilde f(t)$ for $t=0,1,\ldots, T_0-1$.
	Define the multiset $\tilde {\bf f}=[\tilde f(0), \tilde f(1), \ldots, \tilde f(T_0-1)]$.
	
	\noindent 2: Let $f=\tilde f(0)$. As each $f$ has a unique $L$-bit ID, use the 4B5B $M$-symmetrization mapping in \cite{Quasi2018} to map $f$ to an $M$-trit codeword $(w(0),w(1), \ldots, w(M-1))$ with $M=\lceil L /4 \rceil *5+6$.
	
	\noindent 3: Select two primes $P_1 >P_0 \ge \lceil n /(1-p_0) \rceil$.
	
	\noindent 4: For each $s=1,2,\ldots, M-1$, generate independent and uniformly distributed random variables $r_0(s) \in [1,P_0-1]$, $r_1(s) \in [1,P_1-1]$, $b_0(s) \in [0,P_0-1]$ and $b_1(s) \in [0,P_1-1]$.
	
	\noindent 5: For each $t$, compute the following variables:
	
	\noindent 6: $q=\lfloor t/M\rfloor$, $s =(t \;\mbox{mod}\; M)$.
	
	\noindent 7: If $w(s)=2$, let  $f(t)=f$.
	
	\noindent 8: If $w(s) = 1$, let $f(t)$ denote the output channel from the multiset-enhanced modular clock algorithm in Algorithm \ref{alg:clock},  
	with the period $P_1$, the slope $r_1(s)$, the bias $b_1(s)$, the time index $q$, and the multiset $\tilde{f}$.
	
	\noindent 9: If $w(s) = 0$, let $f(t)$ denote the output channel from the multiset-enhanced modular clock algorithm in Algorithm \ref{alg:clock},  
	with the period $P_0$, the slope $r_0(s)$, the bias $b_0(s)$, the time index $q$, and the multiset $\tilde{f}$.
\end{algorithm}

To construct the  QR-LC-LSH4 CH sequence, we combine the multiset-enhanced  modular clock algorithm in Algorithm \ref{alg:clock} and the $M$-trit codeword $(w(0), w(1), \ldots, w(M-1))$ obtained from the strong ternary symmetrization mapping in \cite{Quasi2018}.
Specifically, we group every $M$ consecutive time slots in a frame and have a user to play role $w(s)$
in the $s^{th}$ time slot in a frame. As there are three roles from the ternary mapping, there are three sequences for each user: the 0-sequence, the 1-sequence, and the 2-sequence.
For user $i$, we select two {\em primes} $P_{i,0}$ and $P_{i,1}$ such that $\lceil n_i/(1-p_0) \rceil \le P_{i,0} < P_{i,1}$. The parameter $p_0$ acts as the probability to select a channel from the multiset.
A 0-sequence (resp. 1-sequence) of user $i$ is then constructed by using the multiset-enhanced modular clock algorithm with the prime $P_{i,0}$ (resp. $P_{i,1}$).
The slope parameter and the bias parameter are selected at random.
A 2-sequence is a ``stay'' sequence in which the ID channel is used in every time slot. Then the CH sequence of a user is constructed by interleaving $M$ $\{0/1/2\}$-sequences according to its $M$-trit codeword,  i.e.,
using a $w(s)$-sequence  in the time slots $s, s+M, s+2M, \ldots$, for $s=0,1,\ldots, M-1$.
Let $\{f_i(t), t\ge 0\}$ be the CH sequence for user $i$, $i=1$ and 2.
The insight behind our construction is that the two users will rendezvous immediately at time 0 during the 2-sequence if  both users select the same channel as their IDs
and their clocks are synchronized. On the other hand, either their clocks are not synchronized or their IDs are different, the strong ternary symmetrization mapping in \cite{Quasi2018}
ensures that there exists some time $\tau$ such that the subsequence $\{f_1(\tau),f_1(\tau+M), f_1(\tau+2M), \ldots \}$
and the subsequence $\{f_2(\tau),f_2(\tau+M), f_2(\tau+2M), \ldots \}$
are generated by the multiset-enhanced modular clock algorithm with two different primes. These two users are then guaranteed to rendezvous from the Chinese Remainder Theorem for the multiset-enhanced modular clock algorithm.
This leads to the following theorem.

\bthe{MTTRb}
Consider the sym/async/hetero/ID MRP.
Suppose that each channel is uniquely represented by an $L$-bit ID.
If the two users follow the QR-LC-LSH4 algorithm in Algorithm \ref{alg:QR}, then they are guaranteed to rendezvous within $MP_{1,1}P_{2,1}$ time slots, where $M = \left\lceil \frac{L}{4} \right\rceil \times 5 + 6$.
\ethe

\bproof  
The main difference between the QR-LC-LSH4 algorithm and the QR algorithm in \cite{Quasi2018} lies in the  
channel replacement policy in Step 3 of the modular clock algorithm \cite{Quasi2018}.  
Note that the QR-LC-LSH4 algorithm replaces a channel by selecting it at random from the {\em multiset}.  
In contrast, the QR algorithm replaces a channel by selecting it at random from the {\em available channel set}.  
The result in \rthe{MTTRb} then follows directly from the MTTR bound in Theorem 4 of \cite{Quasi2018}.  
\eproof  

As the MTTR of the QR algorithm is $O((\log N)n_1 n_2)$, the MTTR of the QR-LC-LSH4 algorithm is also $O((\log N)n_1 n_2)$. Compared to the random algorithm, it is expected that our QR-LC-LSH4 algorithm not only has a bounded MTTR but also a lower ETTR when the Jaccard index of the available channel sets of two users is high.

\section{Simulations}
\label{sec:sim}

Our simulation setup follows the procedure outlined in \cite{LSH} for generating the available channel sets of the two users. In each run, we record the TTR. A total of 10,000 experiments are conducted, and the ETTR is estimated by averaging the TTRs across all trials. To compute the (measured) MTTR, we partition the 10,000 experiments into 100 batches, each containing 100 trials. The maximum TTR from each batch is extracted, and the (measured) MTTR is obtained by averaging these 100 maximum values.

\bsubsec{Numerical results for the LC-LSH algorithms}{numlclsh}

In this section, we set the number of time slots to 10,000 and compare our LC-LSH algorithm, described in Algorithm \ref{alg:LCLSH}, with the random algorithm and the LSH2 algorithm \cite{LSH} in the synchronous setting. 
Since the LSH2 algorithm requires a global channel enumeration system, the comparison assumes the existence of such a system, with channels (frequencies) indexed from 0 to $N-1$, where $N$ is the total number of channels. 
In our simulations, the ID of channel $i$ is simply the binary representation of $i$.

\begin{figure}[ht]
	\centering
	\includegraphics[width=0.45\textwidth]{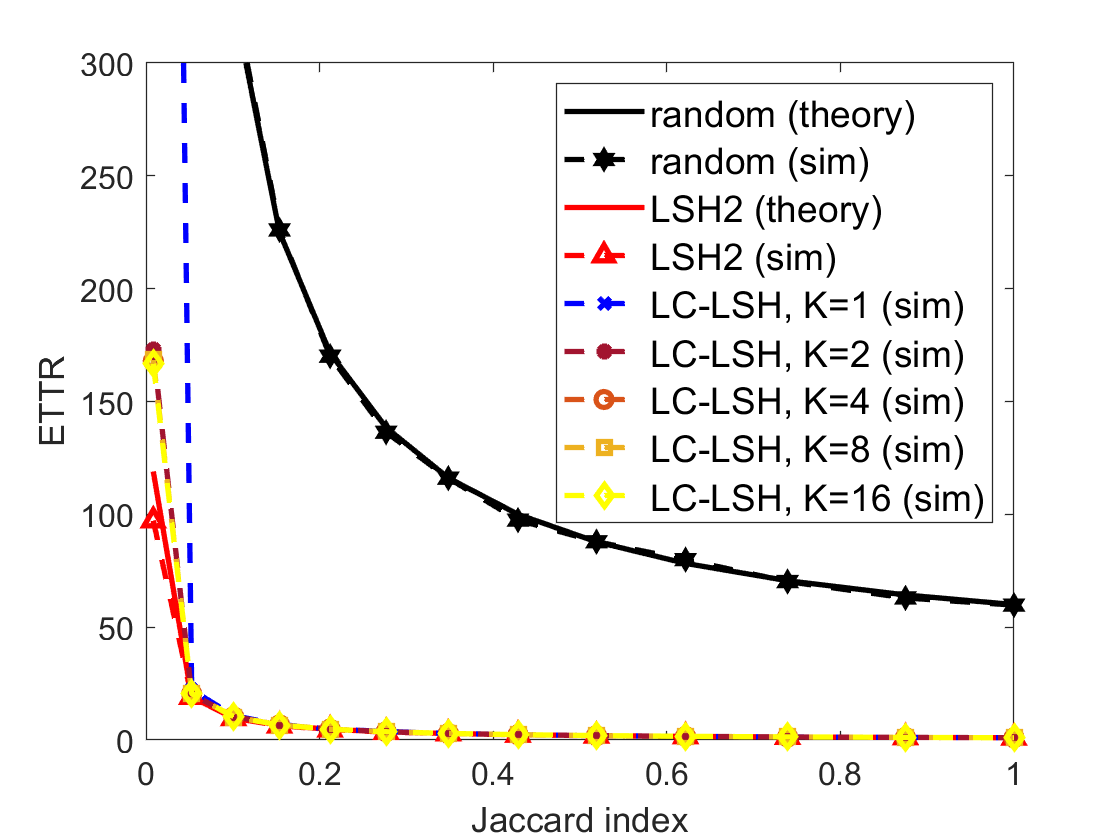}
	\caption{The ETTRs of the LC-LSH algorithm in the synchronous setting with $N=256$, $n_{1}=n_{2}=60$.}
	\label{fig:comparelclshETTR}
\end{figure}

In \rfig{comparelclshETTR}, we show the ETTRs of the {\color{black}LC}-LSH algorithm as a function of the Jaccard index when $N=256$, $n_1=n_2=60$.  We plot the ETTRs for $K=1,2,4,8$ and $16$.
As expected, the simulation results confirm that the ETTRs are very close to the theoretical value $1/J$ when $K \ge 2$. To our surprise, selecting $K \geq 2$ suffices to ensure that the intervals associated with available channels are nearly evenly spaced. This outcome likely stems from the {\em random selection} of available channels in \cite{LSH}.

\begin{figure}[ht]
\centering
\includegraphics[width=0.45\textwidth]{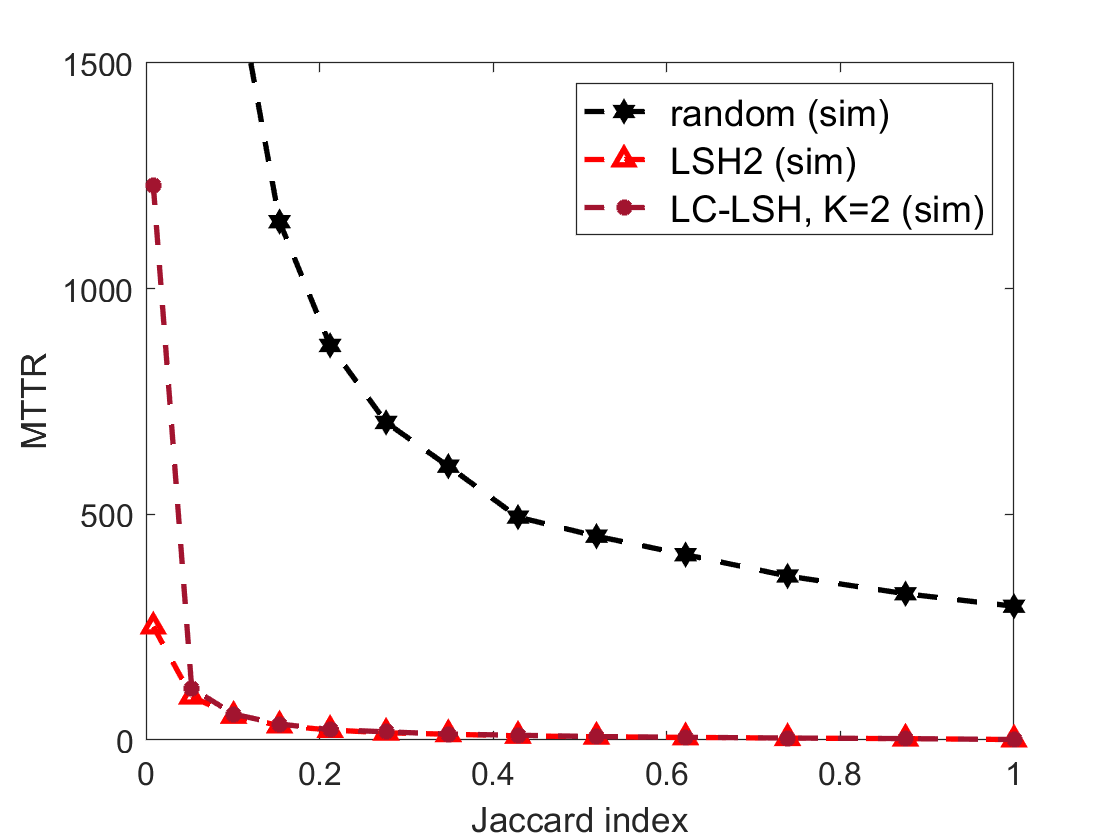}
\caption{The MTTR of the LC-LSH algorithm in the synchronous setting with $N=256$, $n_{1}=n_{2}=60$.}
\label{fig:comparelclshMTTR}
\end{figure}

Different from the LSH2 algorithm, we note that the (measured) MTTR of our LC-LSH algorithm cannot be bounded as it uses sampling with replacement. Consequently, the MTTR measured in our simulations becomes very large when the Jaccard index is small (as shown in \rfig{comparelclshMTTR}).

\begin{figure}[ht]
\centering
\includegraphics[width=0.45\textwidth]{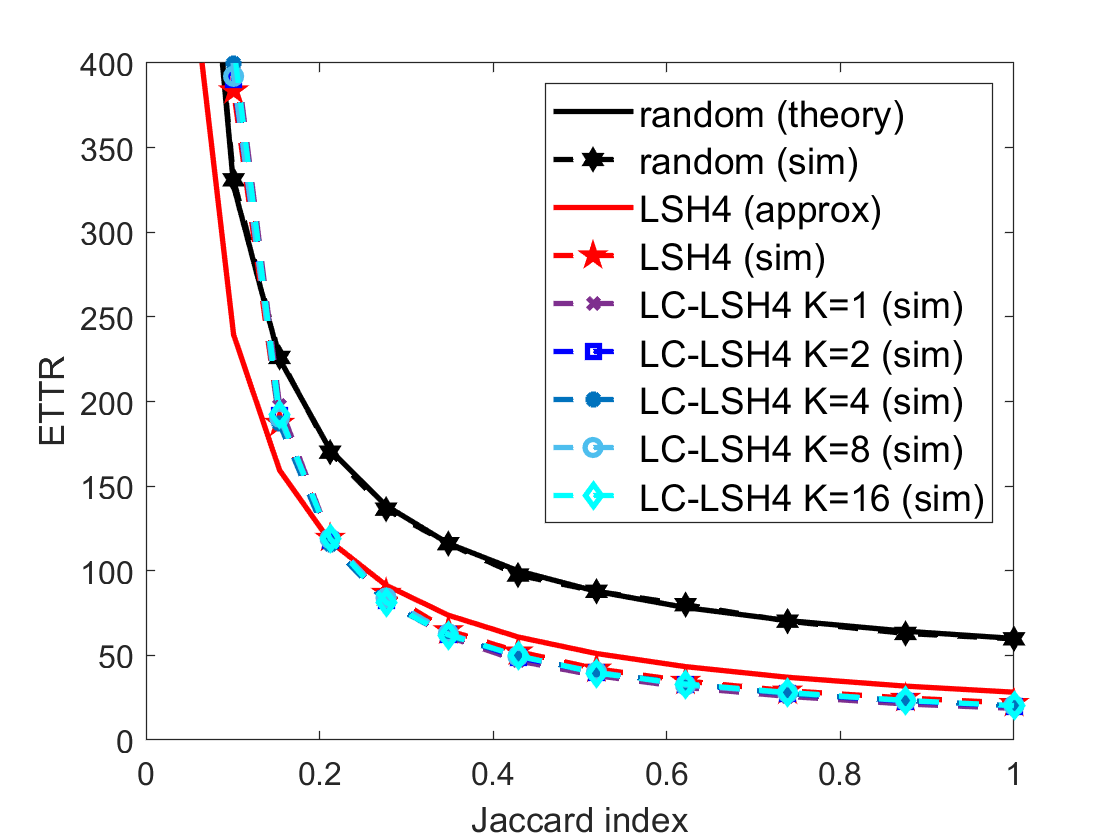}
\caption{The ETTRs of the LC-LSH4 algorithm in the asynchronous setting with $N=256$, $n_{1}=n_{2}=60$.}
\label{fig:comparelclsh4ETTR}
\end{figure}

Then, we compare our LC-LSH4 algorithm in Algorithm \ref{alg:LCLSH4} with the random algorithm and the LSH4 algorithm \cite{LSH} for the asynchronous setting. 
In \rfig{comparelclsh4ETTR}, we compare the ETTRs with $N=256$, $n_{1}=n_{2}=60$. For the LSH4 algorithm and the LC-LSH4 algorithm, we consider the parameter setting with $T_0=20$ and $p_0=0.75$.

\begin{figure}[ht]
\centering
\includegraphics[width=0.45\textwidth]{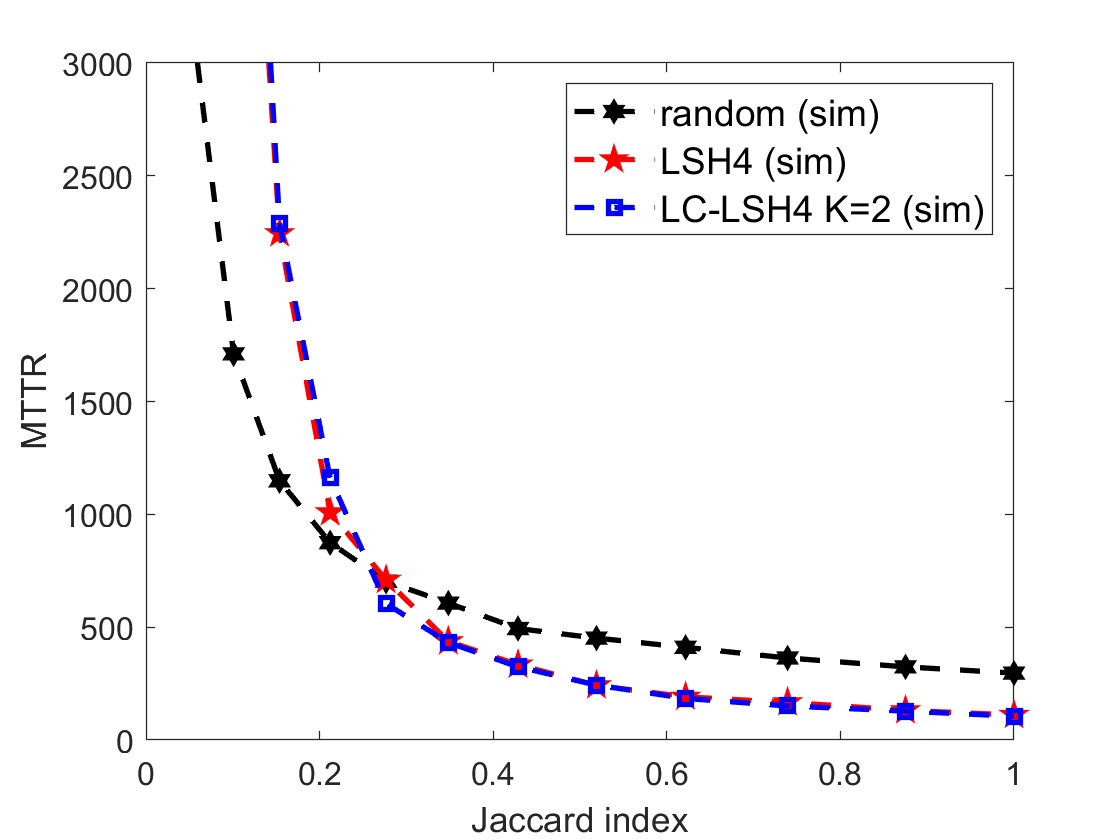}
\caption{The MTTR of the LC-LSH4 algorithm in the asynchronous setting with $N=256$, $n_{1}=n_{2}=60$.}
\label{fig:comparelclsh4MTTR}
\end{figure}

As shown in \rfig{comparelclsh4ETTR}, the ETTRs of the LC-LSH4 algorithm (Algorithm \ref{alg:LCLSH4}) for various values of $K$ are very close to that of the LSH4 algorithm \cite{LSH} and they are significantly better than the random algorithm when the Jaccard index is larger than 0.2. We also note that the simulation results match well with the approximation in \req{LSH41111lc} when the Jaccard index is larger than 0.2. On the other hand, as shown in \rfig{comparelclsh4MTTR}, the (measured) MTTR of our LC-LSH4 algorithm in our simulations is similar to that of the LSH4 algorithm.

\bsubsec{Numerical results for the ASYM-LC-LSH4 algorithm and the QR-LC-LSH4 algorithm  
}{qrlsh4sim}

In this section, we set the number of time slots to be no less than the MTTR bound of each algorithm and compare the performance of our ASYM-LC-LSH4 and QR-LC-LSH4 algorithms with that of the random algorithm, the QR algorithm, and the LC-LSH4 algorithm.

\iffalse
In \rfig{compareqrlclsh4ETTR}, we compare the ETTR of our ASYM-LC-LSH4 algorithm and QR-LC-LSH4 algorithm with the random algorithm, the QR algorithm, and the LC-LSH4 algorithm.
As shown in \rfig{compareqrlclsh4ETTR}, the ETTR of the QR algorithm is comparable to that of the random algorithm, as stated in \cite{Quasi2018}. On the other hand, the ETTR of our proposed ASYM-LC-LSH4 algorithm and QR-LC-LSH4 algorithm is comparable to that of the LC-LSH4 algorithm.
This is because the multiset-enhanced modular clock algorithm (Algorithm \ref{alg:clock}) selects channels in a manner similar to the LC-LSH4 algorithm. Therefore, they behave similarly to the LC-LSH4 algorithm in terms of ETTR.
To be more specific, we know that the clock $k$ {\color{black}in Algorithm \ref{alg:clock}} primarily determines the channel selection of the CH sequence. In Step 1, the value of $k$ ranges from $0$ to $P - 1 = 251 - 1 = 250$. Since $n = 60$, there is a probability of $\frac{60}{251} \approx 0.24$ to select a channel from the available channel set (Step 2 of Algorithm \ref{alg:clock}) and a probability of $1 - \frac{60}{251} \approx 0.76$ to select a channel from the multiset (Step 3 of Algorithm \ref{alg:clock}).
This behavior is similar to the LC-LSH4 algorithm, which has a probability of $p_0 = 0.25$ for selecting a channel from the available channel set and a probability of $0.75$ for selecting a channel from the multiset.
\fi

\begin{figure}[ht]
\centering
\includegraphics[width=0.45\textwidth]{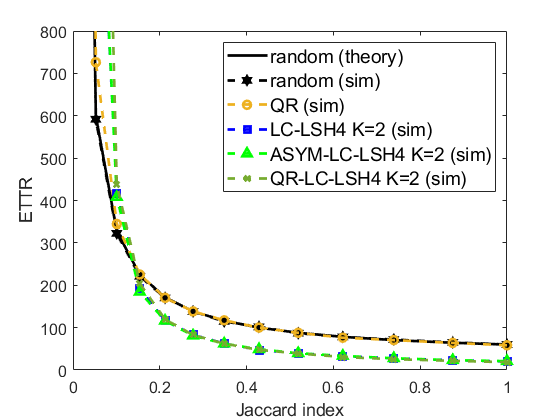}
\caption{The ETTR of the ASYM-LC-LSH4 algorithm and the QR-LC-LSH4 algorithm in the asynchronous setting with $N=256$ and $n_{1}=n_{2}=60$.}
\label{fig:compareqrlclsh4ETTR}
\end{figure}

In \rfig{compareqrlclsh4ETTR}, we compare the ETTR performance of our ASYM-LC-LSH4 and QR-LC-LSH4 algorithms with the random algorithm, the QR algorithm, and the LC-LSH4 algorithm.
As shown in \rfig{compareqrlclsh4ETTR}, the ETTR of the QR algorithm is comparable to that of the random algorithm, consistent with the findings in \cite{Quasi2018}. In contrast, the ETTRs of our proposed ASYM-LC-LSH4 and QR-LC-LSH4 algorithms are similar to that of LC-LSH4. This is because the multiset-enhanced modular clock algorithm (Algorithm~\ref{alg:clock}) selects channels in a way that closely resembles the LC-LSH4 algorithm, resulting in similar ETTR behavior.

\begin{figure}[ht]
\centering
\includegraphics[width=0.45\textwidth]{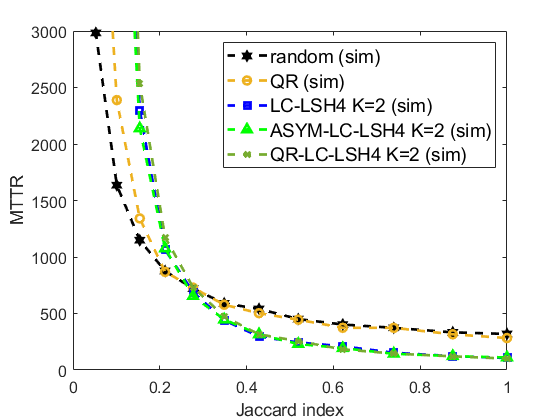}
\caption{The MTTR of the ASYM-LC-LSH4 algorithm and the QR-LC-LSH4 algorithm in the asynchronous setting with $N=256$ and $n_{1}=n_{2}=60$.}
\label{fig:compareqrlclsh4MTTR}
\end{figure}

In \rfig{compareqrlclsh4MTTR}, we present the (measured) MTTR of our proposed ASYM-LC-LSH4 and QR-LC-LSH4 algorithms. When \(J \geq 0.27\), the MTTR of the QR algorithm is comparable to that of the random algorithm. In contrast, the MTTRs of our ASYM-LC-LSH4 and QR-LC-LSH4 algorithms are similar to that of the LC-LSH4 algorithm. Notably, both algorithms outperform the random and QR algorithms when \(J\) is large.
However, when the number of common channels is small (\(J \leq 0.27\), corresponding to \(n_{1,2} \leq 25\)), the MTTRs of ASYM-LC-LSH4 and QR-LC-LSH4 become worse than that of the QR algorithm, which itself performs worse than the random algorithm in this regime. These simulation results indicate that when the overlap in channel availability is limited and no common channels appear in the multiset, embedding the multiset into the modular clock algorithm \cite{Theis2011} does not enhance rendezvous performance. Nevertheless, as noted in \cite{LSH}, users in practical IoT scenarios typically have highly similar available channel sets due to physical proximity--i.e., \(J\) is generally large. Therefore, our ASYM-LC-LSH4 and QR-LC-LSH4 algorithms are well-suited for real-world implementation in such environments.

\section{Conclusion}
\label{sec:con}

\iffalse
In conclusion, this paper introduces a novel LC-LSH algorithm for the multichannel rendezvous problem in IoT networks, eliminating the need for global channel enumeration. By leveraging principles of consistent hashing, the algorithm not only significantly reduces implementation complexity but also achieves ETTRs and MTTRs comparable to those in \cite{LSH}, in both synchronous and asynchronous settings.

To ensure a bounded MTTR in the asynchronous setting, we first employ the multiset-enhanced modular clock algorithm and propose the ASYM-LC-LSH4 algorithm. This algorithm possesses an MTTR bound while keeping its ETTR comparable to that of the LSH algorithm.  
Next, we propose the QR-LC-LSH4 algorithm, which combines our LC-LSH algorithm with the QR algorithm \cite{Quasi2018}. This algorithm inherits the MTTR upper bound of the QR algorithm while achieving an ETTR comparable to that of the LSH algorithm.

These advancements offer a promising direction for future research in neighbor discovery protocols within IoT applications.
\fi

In this paper, we addressed the MRP under the realistic assumption that a global channel enumeration system may be unavailable. By modeling each channel as an $L$-bit identifier (ID), we proposed a suite of low-complexity channel hopping algorithms based on locality-sensitive hashing (LSH), eliminating the need for global channel indexing.

For the synchronous setting, we developed the LC-LSH algorithm, which reduces computational overhead by expanding the hash space and mapping each frequency to multiple virtual frequencies. For the asynchronous setting, we extended this approach and proposed the LC-LSH4 algorithm using a dimensionality reduction technique to maintain low ETTR.

To ensure bounded MTTR, we further introduced the ASYM-LC-LSH4 and QR-LC-LSH4 algorithms by embedding the multiset into the modular clock and quasi-random frameworks, respectively. These algorithms guarantee bounded MTTR while preserving the ETTR advantages of the original LC-LSH4 design.

Extensive simulations demonstrated that the proposed algorithms achieve ETTR and MTTR performance comparable to or better than existing algorithms, especially in scenarios where the available channel sets are similar. These results suggest that our algorithms are well-suited for practical deployment in asynchronous and heterogeneous IoT environments without relying on a globally consistent channel labeling scheme.

\iffalse
In conclusion, this paper introduces a novel LC-LSH algorithm for the multichannel rendezvous problem in IoT networks, eliminating the need for global channel enumeration. By leveraging principles of consistent hashing, the algorithm not only significantly reduces implementation complexity but also achieves ETTRs and MTTRs comparable to those in \cite{LSH}, in both synchronous and asynchronous settings.

To ensure a bounded MTTR in the asynchronous setting, we first employ the multiset-enhanced modular clock algorithm and propose the ASYM-LC-LSH4 algorithm. This algorithm possesses an MTTR bound while keeping its ETTR comparable to that of the LSH algorithm.  
Next, we propose the QR-LC-LSH4 algorithm, which combines our LC-LSH algorithm with the QR algorithm \cite{Quasi2018}. This algorithm inherits the MTTR upper bound of the QR algorithm while achieving an ETTR comparable to that of the LSH algorithm.

These advancements offer a promising direction for future research in neighbor discovery protocols within IoT applications.
\fi

\begin{IEEEbiography}[{\includegraphics[width=1in,height=1.25in,clip,keepaspectratio]{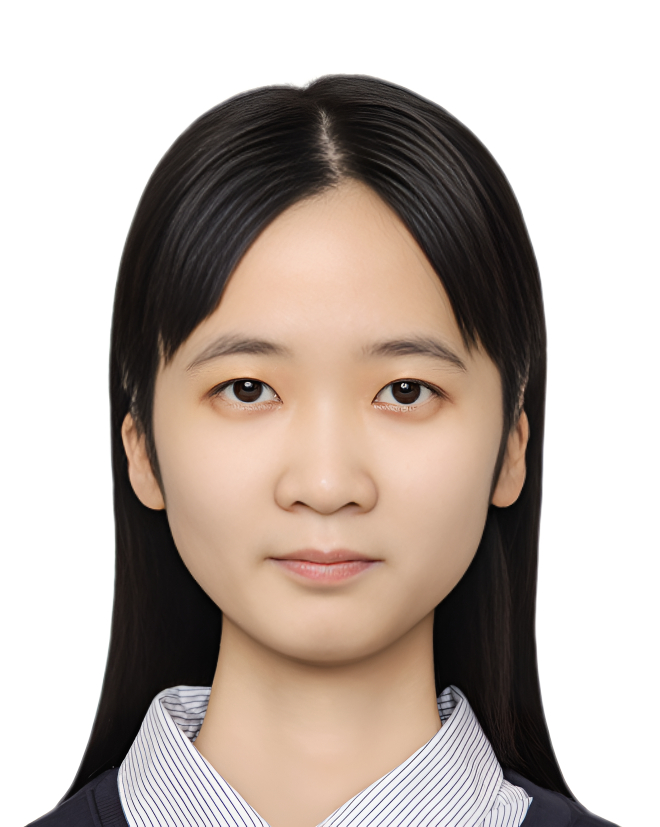}}]
{Yi-Chia Cheng}
received the B.S. degree in Communications, Navigation and Control Engineering from National Taiwan Ocean University, Keelung, Taiwan, in 2023, and the M.S. degree from the Institute of Communications Engineering, National Tsing Hua University, Hsinchu, Taiwan, in 2025. Her research interests include 5G and beyond wireless communication technologies.
\end{IEEEbiography}

\begin{IEEEbiography}[{\includegraphics[width=1in,height=1.25in,clip,keepaspectratio]{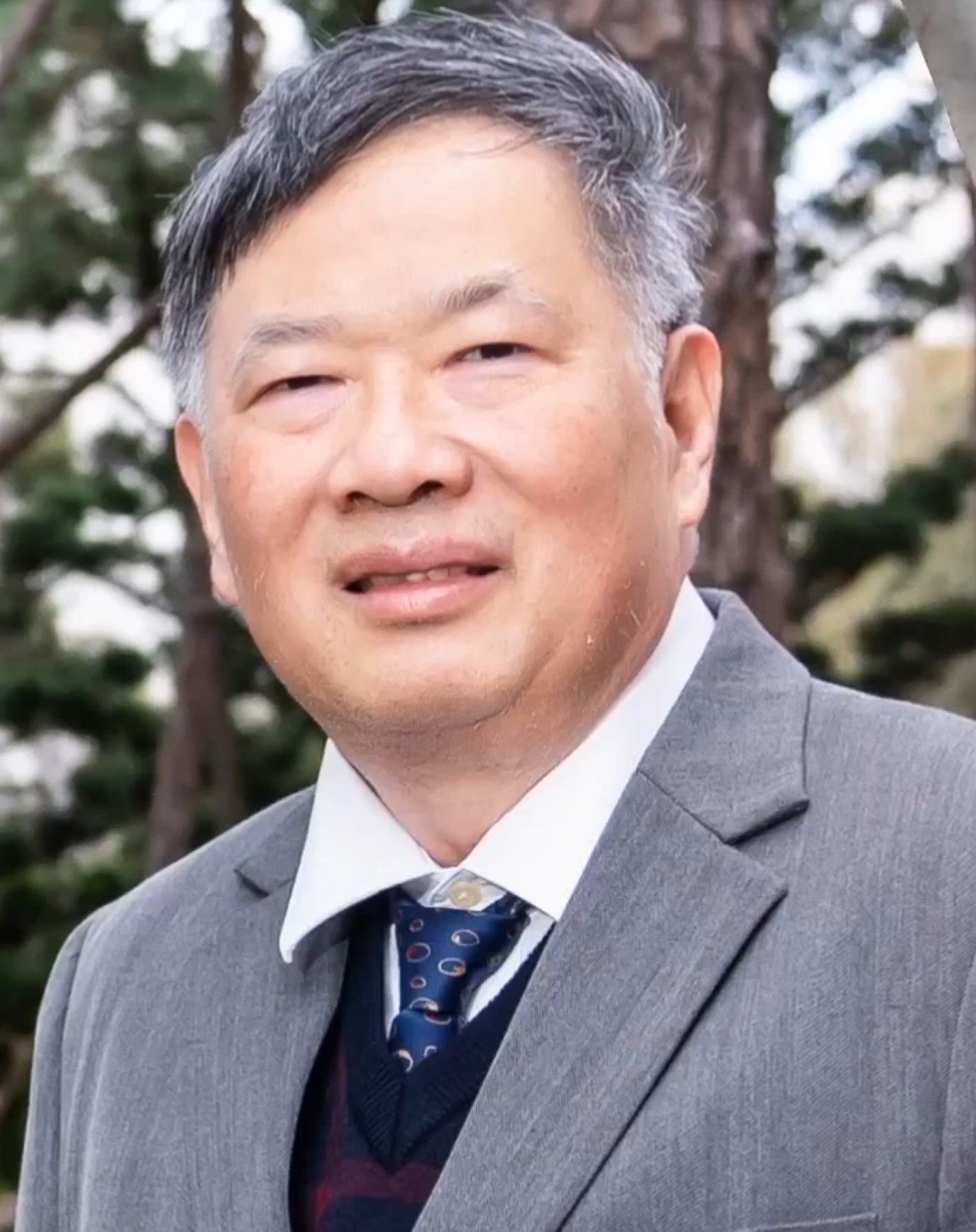}}]
	{Cheng-Shang Chang}
	(S'85-M'86-M'89-SM'93-F'04)
	received the B.S. degree from National Taiwan
	University, Taipei, Taiwan, in 1983, and the M.S.
	and Ph.D. degrees from Columbia University, New
	York, NY, USA, in 1986 and 1989, respectively, all
	in electrical engineering.
	
	From 1989 to 1993, he was employed as a
	Research Staff Member with the IBM Thomas J.
	Watson Research Center, Yorktown Heights, NY,
	USA. Since 1993, he has been with the Department
	of Electrical Engineering, National Tsing Hua
	University, Taiwan, where he is a Tsing Hua Distinguished Chair Professor. He is the author
	of the book Performance Guarantees in Communication Networks (Springer,
	2000) and the coauthor of the book Principles, Architectures and Mathematical
	Theory of High Performance Packet Switches (Ministry of Education, R.O.C.,
	2006). His current research interests are concerned with network science, big data analytics,
	mathematical modeling of the Internet, and high-speed switching.
	
	Dr. Chang served as an Editor for {\em OPERATIONS RESEARCH} from 1992 to 1999,
	an Editor for the {\em IEEE/ACM TRANSACTIONS ON NETWORKING} from 2007
	to 2009, and an Editor for the {\em IEEE TRANSACTIONS
		ON NETWORK SCIENCE AND ENGINEERING} from 2014 to 2017. He is currently serving as an Editor-at-Large for the {\em IEEE/ACM
		TRANSACTIONS ON NETWORKING}. He is a member of IFIP Working
	Group 7.3. He received an IBM Outstanding Innovation Award in 1992, an
	IBM Faculty Partnership Award in 2001, and Outstanding Research Awards
	from the National Science Council, Taiwan, in 1998, 2000, and 2002, respectively.
	He also received Outstanding Teaching Awards from both the College
	of EECS and the university itself in 2003. He was appointed as the first Y. Z.
	Hsu Scientific Chair Professor in 2002. He received the Merit NSC Research Fellow Award from the
	National Science Council, R.O.C. in 2011. He also received the Academic Award in 2011 and the National Chair Professorship in 2017 and 2023 from
	the Ministry of Education, R.O.C. He is the recipient of the 2017 IEEE INFOCOM Achievement Award.
\end{IEEEbiography}

\begin{thebibliography}{99}

\bibitem{wocc2024} Y.-C. Cheng and C.-S. Chang,  ``On the multichannel rendezvous problem without global channel enumeration,'' in {\em Proc. IEEE WOCC}, 2024.

\bibitem{Theis2011} N. C. Theis,  R. W. Thomas, and
L. A. DaSilva,  ``Rendezvous for cognitive radios,'' {\em IEEE Transactions on Mobile Computing}, vol. 10, no. 2, pp. 216--227, 2011.

\bibitem{Bian2013} K.  Bian and J.-M. Park, ``Maximizing rendezvous diversity in rendezvous protocols for decentralized cognitive radio networks,'' {\em  IEEE Transactions on Mobile Computing}, vol. 12, no. 7, pp. 1294-1307, 2013.


\bibitem{Book} Z. Gu, Y. Wang, Q.-S. Hua, and F. C. M. Lau, {\em Rendezvous in Distributed Systems:
	Theory, Algorithms and Applications}. Springer, 2017.


\bibitem{GAP2019} C.-S. Chang, J.-P. Sheu, and Y.-J. Lin, ``On the theoretical gap of channel hopping sequences with maximum rendezvous diversity in the multichannel rendezvous problem, {\em IEEE/ACM Transactions on Networking}, vol. 29, no. 4, pp. 1620--1633, Aug. 2021.
%   {\em arXiv preprint}, arXiv:1908.00198, 2019.

\bibitem{ToN2017} C.-S. Chang,  C.-Y. Chen, D.-S. Lee, and W. Liao,
``Efficient encoding of user IDs for nearly optimal expected time-to-rendezvous in heterogeneous cognitive radio networks,'' {\em  IEEE/ACM Transactions on Networking},  vol. 25, no. 6, pp. 3323--3337, 2017.





\bibitem{Chen14} S. Chen, A. Russell, A. Samanta, and R. Sundaram, ``Deterministic blind rendezvous in cognitive radio networks.'' {\em  IEEE 34th International Conference on Distributed Computing Systems (ICDCS)}, pp. 358--367, 2014.


\bibitem{Improved2015}
Z. Gu,  H. Pu, Q.-S. Hua, and F. C. M. Lau, ``Improved rendezvous algorithms for heterogeneous cognitive radio networks,'' In {\em Proc. IEEE INFOCOM}, pp. 154--162, 2015.

\bibitem{Chang18} Y.-C. Chang, C.-S. Chang, and J.-P. Sheu, ``An enhanced fast multi-radio rendezvous algorithm in heterogeneous cognitive
radio networks,'' {\em IEEE Transactions on Cognitive Communications and Networking}, vol. 4, no. 4, pp. 847--859, 2018.

\bibitem{gu2020heterogeneous}
Z.~Gu, Y.~Wang, T.~Shen, and F.~C. Lau, ``On heterogeneous sensing capability
  for distributed rendezvous in cognitive radio networks,'' \emph{IEEE
  Transactions on Mobile Computing}, vol.~20, no.~11, pp. 3211--3226, 2021.
  
\bibitem{LSH} G.-Y. Jiang and C.-S. Chang, ``Locality-sensitive hashing for efficient
  rendezvous search: A new approach,'' \emph{IEEE Transactions on
  Communications}, vol.~72, no.~9, pp. 5674--5687, 2024.

\bibitem{Leskovec2020} J. Leskovec, A. Rajaraman,  and J. D. Ullman, {\em Mining of massive data sets}. Cambridge university press, 2020.

\bibitem{CHashing}    D. Karger, E. Lehman, T. Leighton, M. Levine, D. Lewin, and R. Panigrahy, ``Consistent
hashing and random trees: Distributed caching protocols for relieving hot spots on the
world wide web,'' {\em Proceedings of the Twenty-ninth Annual ACM Symposium on Theory
	of Computing (STOC)}, pp. 654--663, 1997.

\bibitem{KM04} D. R. Karger, and M. Ruhl, ``Simple efficient load balancing algorithms for peer-to-peer systems,''  {\em Proceedings of the sixteenth annual ACM symposium on Parallelism in algorithms and architectures}, pp. 36--43. 2004.

\bibitem{Quasi2018} C.-S. Chang, Y.-C. Chang, and J.-P. Sheu, ``A quasi-random algorithm for anonymous
rendezvous in heterogeneous cognitive radio
networks,'' {\em  arXiv preprint arXiv:1902.06933}, 2019.

\bibitem{CRSEQ}	J. Shin, D. Yang, and C. Kim, ``A channel rendezvous scheme for cognitive radio networks,'' {\em IEEE Communications Letters}, vol. 14, no. 10, pp. 954-956, 2010.

\bibitem{JS2011} Z. Lin, H. Liu, X. Chu, and Y.-W. Leung, ``Jump-stay based channel-hopping algorithm with guaranteed rendezvous for cognitive radio networks,'' in {\em Proc. IEEE INFOCOM 2011}.

\bibitem{DRDS13} Z. Gu, Q.-S. Hua, Y. Wang, and F. C. M. Lau, ``Nearly optimal asynchronous blind
rendezvous algorithm for cognitive radio networks,'' in {\em Proc. IEEE SECON}, 2013.


\bibitem{Matrix2015} G.-Y. Chang, J.-F. Huang,  and Y.-S. Wang, ``Matrix-based channel hopping algorithms for cognitive radio networks,'' {\em IEEE Transactions on Wireless Communications}, vol. 14, no. 5, pp. 2755--2768, 2015.


\bibitem{DSCR2016} B. Yang, M. Zheng, and W. Liang, ``A time-efficient rendezvous algorithm with a full rendezvous degree for heterogeneous cognitive radio networks,'' in {\em Proc. IEEE INFOCOM}, pp. 1--9, 2016.


\bibitem{Wang2022} Y.-H. Wang, G.-C. Yang, and Wing C. Kwong, ``Asynchronous channel-hopping sequences with maximum rendezvous diversity and asymptotic optimal period for cognitive-radio wireless networks,'' {\em IEEE Transactions on Communications}, vol. 70, no. 9, pp. 5853--5866, 2022.
    
\bibitem{QECH}
Z.~Zhang, B.~Yang, M.~Liu, Z.~Li, and X.~Guo, ``A quaternary-encoding-based
  channel hopping algorithm for blind rendezvous in distributed iots,''
  \emph{IEEE Transactions on Communications}, vol.~67, no.~10, pp. 7316--7330,
  2019.


\bibitem{Alpern95} S. Alpern, ``The rendezvous search problem,'' {\em SIAM J. Control Optim.}, vol. 33, pp. 673--683, 1995.

\bibitem{prime} M. El Bachraoui, ``Primes in the interval [2n, 3n],'' {\em Int. J. Contemp. Math. Sci.}, vol. 1, no. 13, pp. 617–621, 2006.



    
\end{thebibliography}
\end{document}